\documentclass[11pt,letterpaper]{article}

\usepackage[includeheadfoot,
            marginratio={1:1,2:3}, 
            width=412pt,
            height=688pt,]{geometry}
            \usepackage{tikz}
\usepackage[compat=1.1.0]{tikz-feynman}
\usepackage[utf8]{inputenc}
\usepackage{amsmath}
\usepackage{amsfonts}
\usepackage{amssymb}
\usepackage{graphicx}
\usepackage{booktabs}
\usepackage{empheq}
\usepackage{paralist}
\usepackage{cite}
\usepackage[hidelinks]{hyperref}
\usepackage{xcolor}
\usepackage{tensor}
\usepackage{adjustbox}
\usepackage{subcaption}
\usepackage{braket}
\usepackage{tikz}
\usepackage{tikz-cd}
\usepackage{yfonts}
\usepackage{enumitem}
\usepackage{physics}
\usepackage{slashed}
\usepackage{mdframed}

\usetikzlibrary{decorations.pathreplacing,calc}
\newcommand{\nc}{\newcommand}
\nc{\lb}{\llbracket}
\nc{\rb}{\rrbracket}
\nc{\gl}{\llbracket}
\nc{\gr}{\rrbracket}
\nc{\bbR}{\mathbb{R}}
\nc{\bbC}{\mathbb{C}}
\nc{\bbZ}{\mathbb{Z}}
\nc{\cO}{\mathcal{O}}
\nc{\cS}{\mathcal{S}}
\nc{\cM}{\mathcal{M}}
\nc{\cT}{\mathcal{T}}
\nc{\cX}{\mathcal{X}}
\nc{\cQ}{\mathcal{Q}}
\nc{\cA}{\mathcal{A}}
\nc{\cD}{\mathcal{D}}
\nc{\cL}{\mathcal{L}}
\nc{\cC}{\mathcal{C}}
\nc{\cG}{\mathcal{G}}
\nc{\cF}{\mathcal{F}}
\nc{\cI}{\mathcal{I}}
\nc{\cN}{\mathcal{N}}
\nc{\pd}{\partial}
\nc{\la}{\lambda}

\newcommand\beq{\begin{equation}}
\newcommand\eeq{\end{equation}}
\nc{\del}{\partial}
\nc{\tri}{\hspace{-3.5pt}\vartriangle\hspace{-3.5pt}}
\nc{\blacktri}{\blacktriangle}
\nc{\eq}[1]{\begin{equation}
                     \begin{split} #1 \end{split}
                     \end{equation}}
\nc{\ul}{\underline}
\nc{\ov}{\overline}
\nc{\fa}{\hat}
\nc{\fb}{\MakeUppercase}
\nc{\fc}{\tilde }
\nc{\Lie}{{\cal L}} 
\nc{\lambdabar}{{\mkern0.75mu\mathchar '26\mkern -9.75mu\lambda}}

\newmuskip\pFqmuskip

\newcommand*\pFq[7][8]{
  \begingroup % only local assignments
  \pFqmuskip=#1mu\relax
  \mathchardef\normalcomma=\mathcode`,
  \mathcode`\,=\string"8000
  \begingroup\lccode`\~=`\,
  \lowercase{\endgroup\let~}\pFqcomma
  {}_{#2}{#3}_{#4}{\left[\left.\genfrac..{0pt}{}{#5}{#6}\right|#7\right]}
  \endgroup
}
\newcommand{\pFqcomma}{{\normalcomma}\mskip\pFqmuskip}

\allowdisplaybreaks[2]

\nc{\Unote}[1]{\textcolor{gray}{#1}}

\begin{document}

\vspace*{1.5cm}
\begin{center}
{\LARGE
\textbf{On the Complexity of Quantum Field Theory}}

%\vspace{.6cm}
\end{center}

\vspace{0.35cm}
\begin{center}
 {\bf Thomas W.~Grimm} and
 {\bf Mick van Vliet}
\end{center}

\vspace{.5cm}
\begin{center} 
\vspace{0.25cm} 
\emph{
Institute for Theoretical Physics, Utrecht University,
\\
Princetonplein 5, 3584 CC Utrecht, 
The Netherlands } \\
\end{center}

\vspace{2.5cm}

%%%%%%%%%%%%%%%%%%%%%%%%%%%%%%%%%%%%%%%%%%%%%%%
%%%%%%%%%%%%%%%%%%%%%%%%%%%%%%%%%%%%%%%%%%%%%%%
%%%%%%%%%%%%%%%%%%%%%%%%%%%%%%%%%%%%%%%%%%%%%%%
%%%%%%%%%%%%%%%%%%%%%%%%%%%%%%%%%%%%%%%%%%%%%%%
%%%%%%%%%%%%%%%%%%%%%%%%%%%%%%%%%%%%%%%%%%%%%%%
%%%%%%%%%%%%%%%%%%%%%%%%%%%%%%%%%%%%%%%%%%%%%%%
%%%%%%%%%%%%%%%%%%%%%%%%%%%%%%%%%%%%%%%%%%%%%%%
%%%%%%%%%%%%%%%%%%%%%%%%%%%%%%%%%%%%%%%%%%%%%%%

\begin{abstract}
\noindent
We initiate a study of the complexity of quantum field theories (QFTs) by proposing a measure of information contained in a QFT and its observables. 
We show that from minimal assertions, one is naturally led to measure complexity by two integers, called format and degree, 
which characterize the information content of the functions and domains required to specify a theory or an observable.   
The strength of this proposal is that it applies to any physical quantity, and can therefore be used for analyzing complexities within an individual QFT, as well as studying the entire space of QFTs. We discuss the physical interpretation of our approach in the context of perturbation theory, symmetries, and the renormalization group. Key applications include the detection of complexity reductions in observables, for example due to algebraic relations, and understanding the emergence of simplicity when considering limits.
The mathematical foundations of our constructions lie in the framework of sharp o-minimality, which ensures that the proposed complexity measure exhibits general properties inferred from consistency and universality.
\end{abstract}

\clearpage

\tableofcontents

%%%%%%%%%%%%%%%%%%%%%%%%%%%%%%%%%%%%%%%%%%%%%%%
%%%%%%%%%%%%%%%%%%%%%%%%%%%%%%%%%%%%%%%%%%%%%%%
%%%%%%%%%%%%%%%%%%%%%%%%%%%%%%%%%%%%%%%%%%%%%%%
%%%%%%%%%%%%%%%%%%%%%%%%%%%%%%%%%%%%%%%%%%%%%%%
%%%%%%%%%%%%%%%%%%%%%%%%%%%%%%%%%%%%%%%%%%%%%%%
%%%%%%%%%%%%%%%%%%%%%%%%%%%%%%%%%%%%%%%%%%%%%%%
%%%%%%%%%%%%%%%%%%%%%%%%%%%%%%%%%%%%%%%%%%%%%%%
%%%%%%%%%%%%%%%%%%%%%%%%%%%%%%%%%%%%%%%%%%%%%%%

\newpage

\parskip=.2cm

%%%%%%%%%%%%%%%%%%%%%%%%%%%%%%%%%%%%%%%%%%%%%%%%%%%%%%%%%%%%%%%%
\section{Introduction}
Quantum field theory (QFT) remains one of the cornerstones of theoretical physics, describing an enormous variety of physical settings with remarkable precision. Though the depth of our understanding of QFTs is ever-increasing, a complete classification remains elusive. Ultimately, one would like to have a notion of a space of QFTs, in which QFTs and their observables are systematically classified, organized, and relations among them can be understood \cite{Douglas:2010ic}. An essential part of such a construction is to understand what data fundamentally defines a QFT, and how much information is needed to uniquely specify one. Information-theoretic methods have been widely investigated in the context of QFTs, but currently a precise notion of information content of the data which defines a QFT does not exist. 

Motivated by these ideas, our aim is to quantify the complexity of a QFT. This notion of complexity\footnote{Note that we are not considering complexity in the sense of quantum information theory, which is believed to play a significant role in holography \cite{Susskind:2014rva,Brown:2015bva,Jefferson:2017sdb}. However, as pointed out in \cite{Grimm:2023xqy}, it is expected that there are connections between these different notions of complexity.} should measure the information contained within a theory in a physically meaningful way, thereby providing a quantitative tool for analyzing the structure of the space of QFTs and their observables. To conceive such a notion, one is first led to consider the fundamental data which defines a QFT. Starting from the microscopic description of the theory we could, for example, attempt to measure the information contained in the Lagrangian by quantifying the complexity of the Lagrangian as a function of the fields. Complexity should then capture the number of fields or species of the theory, as well as the intricacy of the interactions. Alternatively, one could take the observables of a QFT as the fundamental starting point, and count the information contained in scattering amplitudes and correlation functions, thereby measuring the complexity of observables as a function of kinematic variables and parameters of the theory.

In both cases, one is naturally led to the fundamental mathematical problem of quantifying the complexity of functions. This is a challenging problem, and evidently impossible for general functions unless some strong finiteness constraints are imposed. For instance, a generic analytic function already requires infinitely many coefficients to be specified in a power series expansion, and thus appears to contain an infinite amount of information. The field of mathematics which studies spaces and functions which contain only a finite amount of information is called tame geometry. Objects in this theory are constrained by a tameness axiom called o-minimality, which essentially imposes that they have finitely many geometric features \cite{VdDries}. Remarkably, it was found that many functions arising in QFTs satisfy this tameness axiom \cite{Grimm:2021vpn,Douglas:2022ynw,Douglas:2023fcg,Grimm:2024hdx}, which suggests that they can indeed be described by a finite amount of information \cite{Grimm:2023xqy,Grimm:2024mbw}.

In this work, we draw inspiration from these ideas to explore how to assign a physically meaningful notion of complexity to a QFT. The complexity of a tame function may be captured by two explicitly computable numbers, called format and degree \cite{binyamini2022sharply,beyondo-min}. These numbers come from a framework called sharp o-minimality and provide a consistent and universal way to measure the complexity of functions and domains. We implement this prescription to the Lagrangians and observables of QFT, and propose this as a definition of \textit{QFT complexity}. Even before considering the intricacies of the precise definition of such a notion, we mention four compelling reasons why we believe that our proposal is promising from the outset: 
\begin{enumerate}
    \item[(i)] It consists of two numbers rather than one, and therefore makes a finer distinction between the number of fields and their interactions;
    \item[(ii)] It is absolute, and does not require a reference QFT to be defined relative to;
    \item[(iii)] It is able to describe many complicated functions that appear in QFTs;
    \item[(iv)] It relies on a recently introduced and remarkably general mathematical notion of complexity with properties based on consistency and universality.
\end{enumerate}

After discussing ideas to define QFT complexity, we investigate the physical meaning of our proposals. We do this by commenting on how it relates to various standard QFT aspects, such as symmetries, perturbation theory, and renormalization group flow. In addition, our framework allows us to compare the complexity of the microscopic description of a theory to the complexity of its observables. Our analysis will be exploratory in nature, initiating the first steps but leaving many questions open for future research.

\paragraph{Outline.} The outline of this work is as follows. In section \ref{QFTactions}, we discuss how to assign a complexity to the action of a QFT, and argue that this naturally leads us to measure the complexity of functions using the concept of format and degree. In particular, we show how format and degree encode the number of degrees of fields and the interactions of a theory. We then give a brief invitation to the underlying mathematical framework of sharp o-minimality, and describe the challenge of assigning a complexity to a large class of functions.
In section \ref{sec:amp}, we apply these complexity principles to the observables of a QFT. We then investigate the physical interpretation of this notion of complexity, by considering various aspects of QFTs such as perturbation theory, symmetries, and renormalization. We conclude in section \ref{sec:concl} and provide an outlook on open questions.

%%%%%%%%%%%%%%%%%%%%%%%%%%%%%%%%%%%%%%%%%%%%%%%%%%%%%%%%%%%%%%%%
\section{Complexity of QFT actions} \label{QFTactions}
The purpose of this section is to explore how to assign a complexity to QFTs, by considering their microscopic description. In particular, we will focus on theories which can be described by quantizing a classical action, so that the data defining the QFT consists of a collection of fields $\Phi$ and a Lagrangian $\cL(\Phi)$. By studying progressively more complicated theories, we are naturally led towards the mathematical framework of sharp o-minimality, in which it is possible to explicitly quantify the complexity of functions and domains. 

%%%%%%%%%%%%%%%%%%%%%%%%%%%%%%%%%%%%%%%%%%%%%%%%%%%%%%%%%%%%%%%%
\subsection{Complexity of actions -- scalar QFTs}

\paragraph{Single scalar with algebraic potential.} Let us start with a simple example, namely a $d$-dimensional theory consisting of a single real scalar field $\phi$ with polynomial self-interactions. The Lagrangian then takes the form 
\begin{equation}
    \cL(\phi) = -\frac{1}{2}\pd_\mu \phi \, \pd^ \mu \phi -V(\phi),
\end{equation}
where $V(\phi)= \lambda_D \phi^ D + \lambda_{D-1}\phi^{D-1}+ \ldots + \lambda_0$ is a polynomial. What is the complexity of this Lagrangian? Assuming that the field $\phi$ is canonically normalized, the Lagrangian is specified completely by the polynomial $V(\phi)$, and the question reduces to measuring the complexity of a polynomial in one variable. A natural candidate is the degree $D$ of $V$, since this Lagrangian requires $D$ real numbers to be uniquely specified, namely the coupling constants $\lambda_D,\ldots,\lambda_0$. The complexity of this set of QFTs would then be indexed by a single integer $D$.

However, this measure is not sufficiently refined to accurately reflect the complexity of a polynomial potential, since it only sees the highest-power interaction term. There may be special algebraic relations among the coefficients $\lambda_0,\ldots,\lambda_D$ which simplify the potential significantly. For instance, consider the algebraic relation $\lambda_0=\cdots=\lambda_{D-1}=0$, so that the potential reduces to a monomial $V(\phi)=\lambda_D\phi^D$. Naturally, the QFT defined by this polynomial potential should have a lower complexity than the one for which all coefficients $\lambda_0,\ldots,\lambda_D$ are free. This example already calls for a more refined notion of complexity, and we will resolve this issue later.

\paragraph{Scalars with algebraic potential.} A natural extension of the previous example is to increase the number of degrees of freedom.\footnote{By number of degrees of freedom, we refer to the number of fields in the theory throughout this paper.} Hence, we now consider a theory with $N$ real canonically normalized scalar fields $\phi_1,\ldots,\phi_N$ described by a Lagrangian
\begin{equation}\label{eq:Lscalars}
    \cL(\phi_1,\ldots,\phi_N) = -\frac{1}{2}\sum_{k=1}^N \pd_\mu \phi_k \, \pd^ \mu \phi_k -V(\phi_1,\ldots,\phi_N)\, .
\end{equation}
The potential $V$ is now taken to be a general polynomial of degree $D$
\begin{equation}
    V(\phi_1,\ldots,\phi_N) = \sum_{I,\,|I|\leq D} \lambda_I \phi^I\, .
\end{equation}
In this notation, $I=(I_1,\ldots,I_N)$ is a multi-index, and we define
\begin{equation}
    |I|=I_1+\cdots+I_N, \quad \lambda_I = \lambda_{I_1\cdots I_N}, \quad \phi^I = \phi_1^{I_1} \cdots \phi_N^{I_N}\, .
\end{equation}

As in the previous example, the complexity of this QFT completely lies in the polynomial potential $V(\phi_1,\ldots,\phi_N)$. In addition to the degree $D$, the complexity of the theory should now also depend on the number of fields $N$. A first natural attempt at  quantifying the complexity of this theory would be to define it as a number $\cC$ which depends on $N$ and $D$, e.g.~the number of independent coupling constants $\lambda_I$:
\begin{equation}
    \cC =  \frac{(N+D)!}{N!D!} \,.
\end{equation}
There is, however, a fundamental problem with this approach. From the perspective of computational complexity, $N$ and $D$ play a radically different role. This difference is captured by Bézout's bound, which states that the number of common zeros of $N$ polynomials of degree $D$ is bounded above by $D^N$. In particular, it implies that the computational complexity of solving polynomial equations satisfies a lower bound depending polynomially on the degree $D$, while depending \textit{exponentially} on the number of variables $N$. With this in mind, the complexity of this QFT should reflect the difference in the dependence on $N$ and $D$.

One way to proceed is as follows: instead of combining $N$ and $D$ into a single number, we disentangle the two and measure the complexity by \textit{two numbers}. At first, this may seem undesirable, since one can then no longer straightforwardly compare the complexities of two different QFTs. In other words, with this proposal there will be no `complexity ordering' on the space of QFTs. However, as we will argue in the remainder of this paper, quantifying complexity by two numbers is actually surprisingly meaningful. Physically, it reflects the fact that a QFT depends on the number of degrees of freedom and the interactions among them in a rather different way. Mathematically, it has deep connections to quantifying complexity in geometry \cite{binyamini2022sharply,beyondo-min}, which we explain in more detail in appendix \ref{app:geom}. For the class of QFTs described by the Lagrangian of equation \eqref{eq:Lscalars}, we could measure the complexity by the pair of integers $(N,D)$. Let us note that, although this is a good starting point, these integers are not refined enough to capture certain details of the QFT, such as the geometry of the field space and the strength of the coupling constants. We discuss these aspects more carefully later.

\paragraph{Scalars with analytic potential.} Let us now generalize the theory described by the Lagrangian of equation \eqref{eq:Lscalars} one step further, and assume that $V(\phi_1,\ldots,\phi_N)$ is a general analytic function of the fields. We then seem to run into an obstacle, since the potential $V$ could now contain an infinite amount of information. For instance, if we consider a vacuum and expand the potential in a power series, we will need infinitely many coefficients $\lambda_I$ to specify the Lagrangian. Therefore, this set of QFTs already contains theories with infinite information content in their description, for which a notion of complexity seems to break down. 

Among this class of QFTs with analytic potentials, we would now like to find subclasses which can be described with only a finite amount of information. The key idea was already alluded to in our discussion of algebraic potentials, namely to consider algebraic relations among the coefficients in the power series expansion. With a sufficiently strong relation among the infinite set of couplings $\{\lambda_I\}$, e.g.~a recursion relation, we may hope to find a description with finite complexity. As indicated earlier, understanding how these relations reduce complexity would also be valuable for refining the complexity of polynomials.

\paragraph{Information of analytic functions -- Pfaffian chains.} There exists a precise method of making the previous discussion precise. The idea is to use differential equations to encode relations among power series coefficients, which provides a way of describing a fairly general class of analytic functions with a finite amount of information. Consider for instance the analytic function $f(x)=e^x$. The power series $f(x)=\sum_{k=0}^\infty a_ k x^ k$ of this function has infinitely many non-zero coefficients, but they satisfy an infinite set of algebraic relations, i.e.~$a_k=a_{k-1}/k$ for all $k$. These relations may be encapsulated by a single logical statement, namely $\dd f /\dd x=f$.

To generalize this, one first needs the concept of a Pfaffian chain; this is a finite sequence of functions $\zeta_1,\ldots,\zeta_r$ of $n$ variables $x_1,\ldots,x_n$ satisfying a system of differential equations of the form
\begin{equation} \label{Pfaff_chain-diffeq}
    \frac{\pd \zeta_i}{\pd x_j} = P_{ij}(x_1,\ldots,x_n,\zeta_1,\ldots,\zeta_i),
\end{equation}
where the $P_{ij}$ are polynomials. In particular, note that these functions form a chain in the sense that the derivatives of $\zeta_i$ only depend on the preceding functions $\zeta_1,\ldots,\zeta_i$ and not on $\zeta_{i+1},\ldots,\zeta_r$. Given such a chain, a \textit{Pfaffian function} is any function of the form 
\begin{equation}\label{Pfaffian-function}
    f(x_1,\ldots,x_n) = P(x_1,\ldots,x_n,\zeta_1,\ldots,\zeta_r),
\end{equation}
where $P$ is a polynomial in the variables $x_j$ and the functions of the chain $\zeta_i$.

The main significance of these functions is that they satisfy remarkable finiteness theorems: there are precise bounds on the complexity of their topological and computational properties \cite{GabVor04}. The amount of information contained in a Pfaffian chain depends on the data needed to define the chain, i.e.~the length of the chain $r$, the number of variables $n$, and the degrees of the polynomials $P_{ij}$ and $P$. Typically, these are combined into a \textit{format} $\cF$ and a \textit{degree} $\cD$, defined by 
\begin{equation} \label{degree-format-Pfaffalg}
    \cF = r+ n, \qquad \cD = \deg P + \sum_{i,j}\deg P_{ij}\, .
\end{equation}
Together, these two numbers characterize the complexity of a Pfaffian function \cite{beyondo-min,binyamini2022sharply,EffCell}. While this simple prescription will be sufficient for now, it is important to note that the format and degree of a function have a profound geometric and logical interpretation. In addition, the assignment of a format and degree to geometric objects generalizes far beyond the Pfaffian setting, and forms an active area of research in mathematics. We discuss this in greater depth in section~\ref{sec:generalframework} and appendix~\ref{app:geom}.

\paragraph{QFT complexity of scalar theories.} The previous discussion now provides a way to describe a large class of QFTs with a finite amount of information, namely those scalar theories for which the potential $V(\phi_1,\ldots,\phi_N)$ can be expressed as a Pfaffian function. In particular, it suggests that we measure QFT complexity in a general scalar theory by assigning a format and degree $(\cF,\cD)$ to the action. The advantages of this prescription are that it can accurately reflect the number of degrees of freedom and the complexity of interactions, and can describe theories whose actions are formulated with complicated transcendental functions.\footnote{When focusing on renormalizable theories, it is often sufficient to consider operators of finite dimension in the Lagrangian, but in effective theories much more complicated functions may arise. For example, in the effective actions coming from string theory compactifications one frequently encounters complicated functions coming from period integrals.} The number of degrees of freedom is then counted by $\cF$, defined as the sum of the number of scalars $N$ and the number of auxiliary functions $r$ appearing in the Lagrangian. The degree $\cD$ encodes the complexity of the interactions and relations among the couplings. To gain some intuition for this prescription, let us discuss a few examples.

\paragraph{Example: polynomial and fewnomial interactions.} We first compare two theories of a single real scalar $\phi$ with Lagrangians $\cL_{\rm mono}$ and $\cL_{\rm poly}$, given by
\begin{equation}\label{eq:Lpoly}
    \cL_{\rm poly} =  -\frac{1}{2}\pd_\mu \phi \, \pd^ \mu \phi -\sum_{k=0}^D\lambda_k \phi^k
     \,, \qquad   
     \cL_{\rm mono} =  -\frac{1}{2}\pd_\mu \phi \, \pd^ \mu \phi -\lambda \phi^D \,.
\end{equation}
Even though these two Lagrangians have the same polynomial degree $D$, the theory described by $\cL_{\rm mono}$ should have a lower complexity, since it only contains a single interaction term. This is indeed discerned by format and degree in the following way. The format and degree of $\cL_{\rm poly}$ is simply given by $(\cF,\cD)=(1,D)$, since it can be represented as a Pfaffian function with a Pfaffian chain of length zero. However, for $\cL_{\rm mono}$ there is a more information-efficient representation as a Pfaffian function, namely by setting
\begin{align}
    \zeta_1(\phi)& = \frac{1}{\phi} \, , &&  \frac{\pd\zeta_1}{\pd \phi}=- \zeta_1^2 \,, \\
    \zeta_2(\phi)&=  \phi ^D  \,,  && \frac{\pd\zeta_2}{\pd \phi}= D\zeta_1\,\zeta_2, \nonumber
\end{align}
so that $\cL_{\rm mono}(\phi)=-\frac{1}{2}\pd_\mu \phi \,\pd^\mu\phi +\lambda \, \zeta_2(\phi)$, which has format and degree $(\cF,\cD)=(1+2,2+2+1)=(3,5)$. At the price of introducing two auxiliary field variables $\zeta_1$ and $\zeta_2$, we have effectively reduced the complexity of the representation and removed the dependence on the power of the interaction $D$. This representation is known as the \textit{fewnomial} representation, and plays an important role in the information theory of Pfaffian functions \cite{Fewnomials}.

\paragraph{Example: cosine potential.} As another instructive example, consider a theory with a real massive scalar field $\phi$ whose field space is the finite interval $(-L,L)$, described by the Lagrangian
\begin{equation}
    \cL_{\rm cos}(\phi) = -\frac{1}{2}\pd_\mu \phi \,\pd^\mu \phi - \frac{1}{2}m^2\phi^2 + \lambda \cos (k\pi \phi/L) \,,
\end{equation}
where $k$ is a positive integer. Expanded in a power series, the potential would have infinitely many interaction terms. However, this Lagrangian can be expressed by means of Pfaffian functions, using the Pfaffian chain 
\begin{align}
    \zeta_1(\phi)& = \tan(\frac{\pi \phi}{2L})  &&  \frac{\pd\zeta_1}{\pd \phi}=\frac{\pi}{2L}\big(1+ \zeta_1^2 \big) \,, \\
    \zeta_2(\phi)&=  \cos\left(\frac{\pi\phi}{L}\right)    && \frac{\pd\zeta_2}{\pd \phi}= -2\zeta_1\zeta_2\,. \nonumber
\end{align}
Subsequently, we can write $\cL_{\rm cos}$ as a Pfaffian function, 
\begin{equation}
    \cL_{\rm cos}(\phi) = -\frac{1}{2}\pd_\mu \phi \,\pd^\mu \phi - \frac{1}{2}m^2\phi^2 + \lambda \, T_k(\zeta_2(\phi)),
\end{equation}
where $T_k$ is the Chebyshev polynomial of degree $k$. The theory now has a finite complexity, measured by the format and degree $(\cF,\cD)=(1+2,2+2+k)=(3,4+k)$. As a function of $k$, the format of the theory stays fixed, but the degree grows linearly. A curious observation is that the number of minima of the potential, i.e.~the number of vacua, also grows linearly with $k$. This is a first glimpse at a deep geometric fact of Pfaffian functions - their format and degree encode precise bounds for geometric and topological properties of sets constructed from Pfaffian functions, e.g.~the number of zeros of a Pfaffian function. Applied to QFTs, this means that our notion of QFT complexity for actions is able to accurately probe the vacuum structure of the theory. We discuss this further in section \ref{sec:vac}.

\paragraph{Field redefinitions and minimal representations.} The format and degree of a Lagrangian depend on the representation, and a field redefinition may result in an increase or decrease in the complexity. As an example, consider the Hubbard-Stratonovich transformation in $\phi^4$ theory. Starting with the Lagrangian 
\begin{equation}
\cL(\phi)=-\frac{1}{2}\pd_\mu \phi\,\pd^\mu \phi -\lambda \phi^4 \,,
\end{equation}
we introduce an non-dynamical auxiliary scalar $\sigma$ and use the path integral identity
\begin{equation}
     e^{- \lambda \phi^4}  =  \int D\sigma \, e^{-\frac{1}{2}m^2\sigma^2 +\sqrt{2\lambda}\,m\,\sigma\phi^2}\,.
\end{equation}
The resulting transformed Lagrangian is 
\begin{equation}
    \cL(\phi,\sigma) = -  \frac{1}{2}\pd_\mu \phi\,\pd^\mu \phi -\frac{1}{2}m^2\sigma^2 +\sqrt{2\lambda}\,m \,\sigma\phi^2 \,.
\end{equation}
This Lagrangian describes the same QFT, but its format $\cF$ has increased by one and its degree $\cD$ has decreased by one.

The fact that a given QFT can be represented by different Lagrangians allows one to ask, in a quantitative way, whether a given QFT admits a `minimal representation', e.g.~by minimizing the format or the degree. The complexity of this minimal representation would then measure the minimal amount of information needed to specify a theory. As an example, it is well-known that dualities such as AdS/CFT may lead to significantly simpler descriptions of the same theory. A notion of QFT complexity formalizes this idea, and we hope to further explore this question in future research.

%%%%%%%%%%%%%%%%%%%%%%%%%%%%%%%%%%%%%%%%%%%%%%%%%%%%%%%%%%%%%%%%
\subsection{Complexity of actions -- general QFTs} \label{sec:CoA-general}
For theories with only scalar fields, we have found that the format and degree of the Lagrangian, defined by means of a Pfaffian chain, may provide a good measure of QFT complexity. We now extend this discussion to more general theories, including gauge fields, fermions, and non-trivial target spaces. This generalization will lead to a number of conceptual challenges for complexity, and along the way we will discuss how some of these challenges may be resolved.

\paragraph{Abelian gauge theories.} Continuing with bosonic theories for now, let us start by discussing abelian gauge theories. In particular, consider $d$-dimensional $\text{U}(1)$ gauge field $A_\mu$, described by the Lagrangian 
\begin{equation}\label{eq:U1}
    \cL_{\text{U(1)}} = -\frac{1}{4g^2}F_{\mu\nu} F^ {\mu\nu} \, ,
\end{equation}
with $F_{\mu\nu} = \pd_\mu A_{\nu} - \pd_\nu A_\mu$. Compared to Lagrangians with only scalar fields, we encounter three new challenges for complexity: (i) $A_\mu$ is a 1-form instead of a scalar; (ii) there is a gauge symmetry which renders some of the degrees of freedom unphysical; and (iii) the Lagrangian only contains derivatives of the fields.

How should we assign a format and degree to this Lagrangian, with these points in mind? To address point (i), we could treat the components $A_0,\ldots, A_{d-1}$ as $d$ real field variables. For the moment, we thereby disregard non-trivial topological and geometric features of gauge theory, e.g.~in the form of a non-trivial principal bundle for the gauge field. We discuss these later in this section. Among the field components $A_{0},\ldots,A_{d-1}$, only $d-2$ components correspond to physical degrees of freedom. For the purpose of quantifying complexity of the Lagrangian, we note however that all components are needed; the redundancy arising from the gauge symmetry therefore does not reduce the format $\cF$ of the theory. 

\paragraph{Derivative interactions.} The appearance of field derivatives in the action of equation \eqref{eq:U1} is worth discussing more generally. For the scalar theories, we assumed that all fields were canonically normalized and essentially ignored the kinetic term. This was a strong assumption, and more generally the theory could be a sigma model with a non-trivial field space metric. For the gauge theories, we are forced to rely on derivative terms by gauge invariance. Moreover, in effective field theories, one often encounters higher-derivative interaction terms in the Lagrangian.

Since we are now interested in the complexity of the description of the theory, the natural resolution to this issue would be to view the derivatives as new variables of the Lagrangian. That is, instead of viewing $\cL$ as a function of $\Phi$, we include derivatives as variables and view it as a function of $(\Phi,\pd_\mu\Phi,\pd_\mu\pd_\nu\Phi,\ldots)$. For the scalar theories, this allows us to treat the field space metric in the same way as the potential, namely by assigning a format and degree to each component. In the context of abelian gauge theory this would amount to treating the components field strength $F_{\mu\nu}$ as the variables of the Lagrangian; the QFT described by \eqref{eq:U1} would then have complexity $(\cF,\cD)=\big(\tfrac{1}{2}d(d-1),2 \big)$. Taking this viewpoint naturally leads to a dependence of the format $\cF$ on the spacetime dimension $d$.

As a generalization of the previous example, consider a $d$-dimensional $\text{SU}(N)$ gauge theory, with Lagrangian
\begin{equation}\label{eq:SUN}
    \cL_{\text{SU}(N)} = -\frac{1}{4g^2} \tr (F_{\mu\nu} F^ {\mu\nu}) \, ,
\end{equation}
Counting the number of real field variables, one finds that the complexity of this theory is $(\cF,\cD)=\big(\tfrac{1}{2}(N^2-1)d(d-1) , 2 \big) $. Even though the complexity of the theory grows with $N$, it is known that simpler dual descriptions may emerge in large $N$ limits. This phenomenon, where the complexity grows with $N$ but reduces in the strict $N\to\infty$ limit, is the first hint towards an \textit{emergence of simplicity} captured by the complexity of a QFT, which we will encounter several times in this work.

\paragraph{Fermionic theories.} Although our focus lies mainly on bosonic theories, let us nonetheless briefly comment on how these ideas may be implemented for fermionic theories. Since format and degree are based on real-valued variables, the natural way to proceed is to look at the real and imaginary parts of the components of a fermionic field $\psi$. In particular, we view these variables as real variables multiplying a basis of Grassmann numbers. A consequence of the anti-commuting nature of fermions is that there are only finitely many non-vanishing interactions. For instance, a theory of $N$ Dirac fermions $\psi_1,\ldots,\psi_N$ is described by the general Lagrangian
\begin{equation}
    \cL(\psi_1,\ldots,\psi_N) = - \sum_{j=1}^N \overline{\psi}_j i\slashed{\pd} \psi_j - \sum_{k=1}^{D/2}\sum_{j_1,\ldots,j_k'} \lambda_{j_1\cdots j_k'}(\overline\psi_{j_1}\Gamma_{1}\psi_{j'_1}) \cdots (\overline\psi_{j_k}\Gamma_k \psi_{j'_k}) \, .
\end{equation}
The highest-power interaction $D$ depends on the spacetime dimension $d$ and the number of fermions. In terms of format and degree, this means that $\cD$ is bounded by a simple universal function of $\cF$. The complexity of this theory is easier to quantify due to the necessary polynomial nature of a fermionic Lagrangian.

\paragraph{Effective actions from string theory compactifications.} Compactifications of string theory yield an enormous landscape of low-energy effective field theories, and understanding the general features of these theories is an active topic of research called the swampland program \cite{Vafa:2005ui} (see e.g.~\cite{Palti:2019pca} for a review). The effective Lagrangians of these theories often have a geometric origin and are generally built from complicated transcendental functions \cite{Douglas:2006es}. More precisely, they are usually constructed in terms of the period functions of Calabi-Yau manifolds. It was proven in \cite{BKT} that these are tame functions, meaning that they have a finite information content. Subsequently, it was conjectured in \cite{beyondo-min} that these functions admit a well-defined notion of format and degree, generalizing the complexity $(\cF,\cD)$ of Pfaffian functions. Recent progress on this conjecture has been reported in \cite{binyamini2024lognoetherianfunctions}. Establishing this conjecture would imply that one can explicitly measure the complexity of EFTs in the landscape, which further opens the door to using complexity and information theory to study the landscape and the swampland.\footnote{Other information-theoretic approaches to study the landscape and the swampland were considered in \cite{Denef:2006ad,Denef:2017cxt,Stout:2021ubb,Stout:2022phm}.}  We intend to discuss this in greater detail in an upcoming work \cite{GV1}.

\paragraph{Topology and complexity.} So far in our discussion we have mostly dealt with theories with no non-trivial topological features. In general, these features are an important aspect of QFTs, in the form of e.g.~spacetime topology, principal bundles for gauge theories, and target spaces in non-linear sigma models. Because the concept of format and degree which we use in our notion of QFT complexity is ultimately geometric in nature (as explained in more detail in appendix \ref{app:geom}), we believe that they can be generalized to QFTs exhibiting these non-trivial topological features. This may be done by subdividing spaces into trivial local patches and using logical operations to add the local complexities. It would be interesting to explore this further in the future, but we leave such an analysis beyond the scope of this work.

\paragraph{Symmetries.} Let us briefly comment on the role of symmetries in the complexity of the action of QFTs. At an intuitive level, the presence of symmetries may lead to the simplification of a theory, but the way in which this is seen from in the format and degree is subtle. Let us illustrate this by a few examples. The QFT defined by the Lagrangian $\cL(\phi ) = -\frac{1}{2}\pd_\mu \phi \,\pd^\mu \phi  - \lambda \sin (2\pi k \phi) $
initially has infinite complexity, because the potential has infinitely many distinct zeros. However, it also has a global $\bbZ$ symmetry, and after taking a quotient by $\bbZ$ by identifying $\phi\sim \phi+1/k$ we obtain a simple theory with finite complexity. On the other hand, if the theory has a global symmetry group $G$ which acts in a more non-trivial way, then the theory obtained by quotienting by $G$ may have a more complicated geometry and therefore an action of higher complexity. This illustrates that the interaction between symmetries and complexity can be rather complicated.

\subsection{Vacuum structure}\label{sec:vac}
Having discussed how to use format and degree to measure the complexity of a QFT action, let us highlight an application. One of the essential properties of format and degree is that they encode bounds on geometric and topological information. For instance, given a Pfaffian function $f$ with complexity $(\cF,\cD)$, the number of connected components of the solution set to the equation $f=0$ is bounded by \cite{Fewnomials,GabVor04}
\begin{equation} \label{zero-bounds}
    \big|\pi_0(\{x \,|\, f(x)=0\}) \big|  \leq 2^{\cF^2} \cF^\cF \cD ^\cF \,. 
\end{equation}
Generalized definitions of format and degree, to be discussed below and in appendix \ref{app:geom}, satisfy similar bounds growing polynomially in $\cD$ \cite{binyamini2022sharply,beyondo-min}.
When applying this principle to QFTs, we can use this to analyze the vacuum structure of the theory. If the potential $V(\Phi)$ has complexity $(\cF,\cD)$, we can then estimate the number of vacua, i.e.~the number of connected components of the set 
\begin{equation}
    \left\{\Phi \, \Big|\, \frac{\pd V}{\pd \Phi}=0\right\}\,.
\end{equation}
While there is a polynomial growth in $\cD$, these estimates typically grow rapidly with $\cF$, and for simple potentials it is much more effective to count the number of vacua by hand. However, for QFTs with extremely complicated potentials, such as the effective theories arising from string compactifications, these countings are numerically very challenging. In such cases, the bounds in terms of format and degree provide simple and universal estimates for the vacuum structure of the theory.\footnote{This story holds in general for sharp o-minimality, although one has to adjust the bounds depending on the class of functions under consideration. The framework is designed so that the bounds are always explicitly computable in terms of $(\cF,\cD)$.}

An important lesson to draw from these bounds is that the significant computational cost of increasing the format $\cF$ encourages one to find a representation of minimal complexity, i.e.~a formulation of the QFT for which the action has the smallest format.

\subsection{General framework for complexity}\label{sec:generalframework}

In the examples of the previous subsections we have relied on format and degree $(\cF,\cD)$ as defined for Pfaffian functions.  While this idea can be implemented to study complexity aspects for a large class of QFTs, there are many additional aspects that one might want to incorporate. Fortunately, the Pfaffian approach to complexity is part of a far greater framework of complexity for geometric objects, called sharp o-minimality. One of the main goals of this program is to identify classes of functions which satisfy bounds on the number solutions to equations, generalizing the Bézout bound for computational complexity in algebraic geometry. In particular, one assigns an explicitly computable format and degree $(\cF,\cD)$ to these functions, such that the Bézout bound $\cD^\cF$ is replaced by a general polynomial $P_{\cF}(\cD)$ in the degree $\cD$. The format and degree of a function can then be thought of as the complexity of the function. Pfaffian functions provide the first example of such a class of functions, with the corresponding bound given by equation \eqref{zero-bounds}. The development of this theory is an ongoing mathematical program \cite{binyamini2022sharply,beyondo-min}, but recent developments are already able to greatly extend the class of functions for which complexity can be defined, and capture more refined aspects of these functions \cite{binyamini2024lognoetherianfunctions}.

While we will refrain from giving a detailed introduction to sharp o-minimality in main text (see appendix \ref{app:geom} for a brief summary), we do want to highlight some of the main ideas behind the constructions of candidate sharp o-minimal structures and their core properties. The reader who is mainly interested in the applications to QFT, for which the Pfaffian setting often suffices, may safely skip this section.

\paragraph{Local bounds and domain dependence.} The significance of Pfaffian functions lies in the \textit{global} bounds which they satisfy, which makes them behave like polynomials despite being a fairly general class of functions. Generic analytic functions do not satisfy such bounds and may therefore have an infinite complexity. One way to nonetheless include, in addition to Pfaffian functions, more general analytic functions in the framework is to restrict these functions to bounded domains, and to rely on \textit{local} complexity bounds. However, it then becomes evident that the complexity of such a restricted analytic function should depend on the size of the domain. As a simple example, the function $f(x) = \sin(x)$ has infinitely many zeros on the real line, and when restricted to a bounded interval it has a finitely many zeros depending on the size of the interval.

\paragraph{Log-Noetherian functions.} A class of functions which satisfies strong local complexity bounds is given by the log-Noetherian functions introduced in \cite{binyamini2024lognoetherianfunctions}. We will not give the complete definition of these functions (see section 2 of \cite{binyamini2024lognoetherianfunctions}), but rather describe an illustrative case that captures many of their core features. Let us consider a domain of the form 
\begin{equation}\label{eq:domainU}
U =D_\circ(\rho_1)\times \cdots \times D_\circ(\rho_N)  \subset \bbC^N \,,
\end{equation}
where $D_\circ(\rho)=\{z\in\bbC\,|\,|z|<\rho\}$ is the punctured disk of radius $\rho$.
A \textit{log-Noetherian} chain on $U$ consists of bounded holomorphic functions $\xi_1,\ldots,\xi_R: U \to \bbC$ satisfying a system of differential equations
\begin{equation}
    z_j \frac{\pd \xi_i}{\pd z_j} = P_{ij}(\xi_1,\ldots,\xi_R)\, .
\end{equation}
This resembles the definition of a Pfaffian chain, but there are a few crucial differences:
\vspace{-2em}
\begin{enumerate}
    \item[(i)] the triangularity requirement is dropped, and the polynomials $P_{ij}$ may depend on all functions $\xi_1,\ldots,\xi_R$. In this construction, the variables $z_1,\ldots,z_n$ themselves have to be defined as functions in the chain;
    \item[(ii)] the functions and variables are complex-valued, and instead of regular derivatives one uses logarithmic derivatives $z_j\frac{\pd}{\pd z_j}$;
    \item[(iii)] the domain $U$ is bounded and has to be of a specific cellular form, as explained in detail in \cite{binyamini2024lognoetherianfunctions}.\footnote{Essentially, the domains $U$ are constructed inductively by fibering disks and annuli over each other. The radii of these fibered disks and annuli are allowed to vary as a log-Noetherian function on the base space. The domain $U$ given in equation \eqref{eq:domainU} is a simple example of such a domain, in which the radii are constant.}
\end{enumerate}
A log-Noetherian function is then any function $f(z_1,\ldots,z_N) = P(\xi_1,\ldots,\xi_R)$ depending polynomially on the functions in the chain. These functions are known to satisfy local complexity bounds, and it has therefore been conjectured that they are sharply o-minimal. While it is presently not known how to precisely assign a format and degree to such a function and domain, it is suggestive\footnote{To be precise, it is shown in \cite{binyamini2024lognoetherianfunctions} that these functions are \textit{effectively} o-minimal, meaning that they satisfy effectively computable bounds characterized by a single number.} from  \cite{binyamini2024lognoetherianfunctions} that their sum should be
\begin{align}\label{eq:formatLN}
    \cF_{\rm LN} + \cD_{\rm LN}  =  &\ \ R  +\deg P +  \sum_{i,j}\deg P_{ij} \\
    & \ \ + \norm{P} +  \sum_{i,j}  \norm{P_{ij}}  + \max_{i,z\in \overline{U}} |\xi_i (z)| + \cF(U) + \cD(U)\,. \nonumber
\end{align}
Note that the terms in the first line of this equation 
are also present in the case of Pfaffian functions \eqref{degree-format-Pfaffalg}, when taking into account that the variables are now considered as being part of the chain. In the log-Noetherian setting there are now several new terms capturing details of the functions and the domain on which they are defined. In particular, one has the norms $\norm{P}$, $\norm{P_{ij}}$ of the polynomials, which are defined as the sum of the absolute values of their coefficients. These are essential for giving the required local bounds on the function. The domain complexity $\cF(U) + \cD(U)$ given in \cite{binyamini2024lognoetherianfunctions} evaluates to a linear function in $\sum_{j}\rho_j$. In contrast to Pfaffian functions the domain complexity needs to be taken into account as we will illustrate in the following example.

Consider the function $f(z)=\frac{e^{\lambda z}-1}{z}$, with $\lambda$ real and positive, on a punctured disk $D_\circ(\rho)\subseteq\bbC$ of radius $\rho$. This function fits in the log-Noetherian chain given by 
\begin{align}
    \xi_1(z)& = z  &&  z\frac{\pd\xi_1}{\pd z}=  \xi_1   \,, \\
    \xi_2(z)& = e^{ \lambda z}  &&  z\frac{\pd\xi_2}{\pd z}=   \lambda \xi_1\xi_2   \,, \nonumber \\
    \xi_3(z)&= \frac{e^{ \lambda z}-1}{z}     && z\frac{\pd\xi_3}{\pd z}=  \lambda \xi_2-\xi_3  \,. \nonumber
\end{align}
The numbers of zeros of $f$ on $\overline{D_\circ(\rho)}$ is given by $2\lfloor \frac{\lambda\rho}{2\pi} \rfloor$, and this should be reflected by the complexity of the function. Indeed, in the log-Noetherian chain $\lambda$ appears as a coefficient in the polynomials on the right-hand side, and the dependence on $\rho$ appears through the domain complexity $\cF(D_\circ(\rho)) + \cD(D_\circ(\rho))$. This shows that the terms in the second line of equation \eqref{eq:formatLN} are necessary.

\paragraph{Pfaffian closures.} Ultimately, one would like to find the most general class of functions which are sharply o-minimal, and hence admit a meaningful notion of complexity. To this end, after having found a suitable function class $\cS$ which admits a format and degree, a further step which can be taken is to generalize the definition of a Pfaffian chain. One of the key reasons why Pfaffian functions satisfy global complexity bounds such as equation \eqref{zero-bounds}, is that these are inherited from the polynomials which appear in the construction. A natural generalization would then be to replace these polynomials by functions from the class $\cS$. We then consider chains $\zeta_1,\ldots,\zeta_r$ which satisfy 
\begin{equation}
    \frac{\pd \zeta_i}{\pd x_j} = F_{ij}(x_1,\ldots,x_n,\zeta_1,\ldots,\zeta_i)\ ,
\end{equation}
where the $F_{ij}$ belong to some class of functions $\cS$. In analogy to \eqref{Pfaffian-function}  one can now use these functions $\zeta_i$ to introduce functions of the form 
\begin{equation} \label{f-gen}
f(x_1,\ldots,x_n)=F(x_1,\ldots,x_n,\zeta_1,\ldots,\\\zeta_r)\ , 
\end{equation}
where $F$ is a function within the class $\cS$. 
This construction is called the \textit{Pfaffian closure} of $\cS$ \cite{Speisegger99}. It turns out that the construction of the Pfaffian closure gives a novel class of functions that inherit the tameness properties of $\cS$. 

It is now apparent that any definition of a complexity for the functions $f$ in the Pfaffian closure should be formed by combining the complexity of the chain with the complexity of the underlying functions $F_{ij}$ and $F$ appearing in \eqref{f-gen}, including the complexity of their domain $U$. The natural assertion for the format and degree of such a function $f$ is 
\beq
  \cF  = r + \cF_{\cS}(F,\{F_{ij}\},U)\,, \qquad  \cD = \cD_{\cS}(F,\{F_{ij}\},U) \,,
\eeq
where $\cF_{\cS}$ and $\cD_{\cS}$ are some appropriately defined format and degree of the class of functions $\cS$. In this way, the construction is a true generalization of the Pfaffian functions and their format and degree as introduced in equation \eqref{degree-format-Pfaffalg}, by allowing for differential equations with local finiteness properties.

\paragraph{The structure $\bbR_{\rm LN,PF}$.} The proposal of \cite{binyamini2024lognoetherianfunctions} is to combine the two constructions explained above; namely to consider the Pfaffian closure of the log-Noetherian functions. This class of functions, though requiring a somewhat complicated construction, turns out to include a wealth of functions, while still retaining controllable local and global complexity bounds. In fact, it is conjectured that this class of functions is sharply o-minimal as well. Notably, the class includes all algebraic period functions. These are crucial in geometric applications, and also appear ubiquitously in QFTs. For instance, they appear generally in QFT amplitudes through the evaluation of Feynman integrals, and feature strongly in the effective actions obtained from string theory compactifications. We therefore believe that establishing the sharp o-minimality of $\bbR_{\rm LN,PF}$ bears great significance for exploring the complexity of QFTs.

\paragraph{Geometric properties of sharp o-minimality.} An important point to emphasize is that sharp o-minimality is defined not only for functions, but also for sets constructed from these functions. In fact, the latter is more fundamental, since functions can be viewed as sets when considering their graphs. The sets considered in sharp o-minimality are built from solution sets of equations, and include for example the domains considered above. The notion of complexity for solution sets of equations should then extend to any collection of sets generated by performing logical operations, i.e.~taking finitely many unions, intersections, complements, products, and projections. In this way, it is possible to assign a format and degree to these sets. Crucially, the framework imposes certain natural requirements on the behavior of format and degree under these logical operations. As an example, the format and degree of sets $A$ and $B$ obey inequalities\footnote{Actually, as explained in the appendix, format and degree generate a double filtration on the collection of sets; therefore these inequalities need to be interpreted with some care.} such as 
\begin{equation}
    \cD(A\cup B) \leq \cD(A) + \cD(B)\, ,\qquad  \cD(A\cap B) \leq \cD(A) + \cD(B)\, ,
\end{equation}
and
\begin{equation} 
   \cF(A \cup B)  \leq \max(\cF ( A),\cF(B))\, ,\qquad 
   \cF(A \cap B)   \leq \max(\cF ( A),\cF(B)) \, .
\end{equation}
These inequalities indicate that the degree is an additive quantity, whereas the format is maximized. These conditions are motivated by computational complexity theory; the union and intersection represent the `or' and `and' logic gates, so the above inequalities constrain how complexity behaves under these gates. For the interested reader we provide the details in appendix \ref{app:geom}.

%%%%%%%%%%%%%%%%%%%%%%%%%%%%%%%%%%%%%%%%%%%%%%%%%%%%%%%%%%%%%%%%
\section{Complexity of QFT observables}\label{sec:amp}
In the previous section we explored how to assign a complexity to the microscopic description of a QFT, focusing on the information contained in the field content and the Lagrangian. We now change our perspective, and view a QFT through the lens of its observables, focusing mostly on scattering amplitudes and correlation functions. There are three main questions on the complexity of observables which we consider:
\begin{enumerate}
    \item[(i)] What is the complexity of an individual QFT observable? 
    \item[(ii)] What is the complexity of a QFT, based on its observables?
    \item[(iii)] What is the connection between the complexity of the action and the complexity of observables?
    \item[(iv)] How is the complexity of a theory affected by renormalization group flow?
\end{enumerate}
Our aim is not to conclusively answer these questions, but to provide some initial ideas and observations. 

\subsection{Amplitudes and correlation functions}

\paragraph{QFT observables - general aspects.} Consider a general QFT observable $\cO(\kappa,\lambda)$, which we view as a function of kinematic variables (e.g.~positions or momenta) denoted by $\kappa$, and of the parameters of the theory, collectively denoted by $\lambda$. How do we measure the complexity of this object? To answer this question, we will use the same strategy as in the previous section; namely to use functional complexity to assign two numbers $(\cF,\cD)$ to $\cO$ as a function of $\kappa$ and $\lambda$. Before we proceed, let us list the different cases which we should consider. First, we should distinguish whether the observable is obtained explicitly from the quantization of an underlying microscopic theory, or whether it is viewed as an entity on its own. In the first case, we should further discern between perturbative expansions of observables, e.g.~by means of Feynman diagrams, and exact observables, which include all non-perturbative effects. In the second case, where no microscopic description is available, we will analyze the complexity of observables from a bootstrap perspective. We study these case by case below.

\paragraph{Perturbative amplitudes and Feynman integrals.} Let us first consider amplitudes which are obtained perturbatively by means of Feynman diagrams. We will denote such an amplitude by $\cA_{\ell,s}(p,\lambda)$, where $\ell$ is the number of loops, $p=(p_1,\ldots,p_n)$ and $s=(s_1,\ldots,s_n)$ denote the external momenta and spins, and $\lambda$ collectively denotes the couplings of the theory. In perturbation theory, these amplitudes are computed by summing the values of all Feynman diagrams with $\ell$ loops, and the values of the individual diagrams are obtained by performing certain integrals over rational functions known as Feynman integrals. Recently, it was proven that the resulting functions $\cA_{\ell,s}(p,\lambda)$ are tame functions, meaning that they in principle have a finite information content \cite{Douglas:2022ynw}.\footnote{The precise statement is that the amplitudes are definable in the o-minimal structure $\bbR_{\rm an,exp}$, and is based on the definability of the period map \cite{BKT}.} This result opens the door to the functional complexity methods used in the previous subsection. In fact, from \cite{Douglas:2022ynw} it follows that the amplitudes belong to the Pfaffian closure of log-Noetherian functions introduced in section \ref{sec:generalframework}, which is conjectured to be sharply o-minimal in \cite{binyamini2024lognoetherianfunctions}.

The complexity of perturbative amplitudes is expected to grow in at least two ways. Firstly, there is a clear dependence on the number of external particles, since the format counts the number of variables. Secondly, at higher orders in perturbation theory one must sum over a larger set of more complicated diagrams, leading to more complicated Feynman integrals and therefore functions with higher $\cF$ and $\cD$.

\paragraph{Exact observables and simplicity.}
Perturbation theory provides, with the exception of rare cases, only asymptotic expansions to exact observables. While the terms in the perturbative expansion reveal much of the underlying QFT, ultimately one should consider the complexity of exact observables. Fundamentally, the exact observables arise from summing over infinitely many physical processes, and therefore there is initially no reason to expect that these observables can be described with a finite amount of information. However, upon taking infinite limits, remarkable reductions of complexity may appear.

Let us illustrate this by a simple example. Consider the sequence of functions 
\begin{equation}
    f_j(x) =  \sum_{k=0}^j \frac{x^k}{k!} \,.
\end{equation}
In terms of functional complexity, the function $f_j$ has format and degree $(\cF,\cD)=(1,j)$ when it is viewed as a arbitrary polynomial, and in the limit $j\to\infty$, this complexity diverges. However, these polynomials are not arbitrary, and the limiting function is $f(x) = e^x$, which has finite format and degree $(\cF,\cD)=(2,2)$ because it admits a simple Pfaffian chain representation as $\pd f/\pd x=f$. The infinitely many algebraic relations among the coefficients ensure that the information content of this sequence of function does not diverge, such that the limiting function has finite complexity. We call this phenomenon the \textit{emergence of simplicity}.

This example offers some hope that exact QFT observables have a finite complexity, despite the fact that the complexity of the perturbative expansion grows order by order. Let us test this in a simple example. A useful testing ground is $\phi^4$ theory on a point spacetime, where exact non-perturbative calculations of all observables are possible. The Euclidean action of this theory, in a convenient parametrization, is given by  
\begin{equation}
    S[\phi] = g\phi^2 + \frac{g}{8}\phi^4\,,
\end{equation}
and the partition function is given by the ordinary integral
\begin{equation}
    Z(g) = \int \dd \phi \, e^{-S[\phi]}\,.
\end{equation}
Calculating the partition function by means of a perturbative expansion\footnote{The coupling $g$ is proportional to $1/\lambda$, where $\lambda \phi^4/4!$ is the standard way of parametrizing the interaction.} in $1/g$ yields the asymptotic series
\begin{equation}
    Z(g) \sim \sum_{k=0}^\infty  \frac{(-1)^k}{8 ^k \,k!} \Gamma(2k+\tfrac{1}{2}) g^{-k-1/2} \,.
\end{equation}
As a Pfaffian function of $g$, the complexity of the partition function grows linearly with the order of the expansion. However, this power series diverges and is only an asymptotic expansion. Nonetheless, in this case the exact partition function can be obtained from a Borel resummation of the series \cite{Grimm:2024hdx}, which yields
\begin{equation}
    Z(g) = \sqrt{2} \, e^g  K_{1/4}(g) \, ,
\end{equation}
where $K_{1/4}$ is a modified Bessel function. It was shown in \cite{Grimm:2023xqy} that this exact partition function $Z(g)$ can be written as a Pfaffian function with complexity $(\cF,\cD)=(4,7)$. Thus, the diverging information content of the perturbative partition function can be tamed, and we observe an emergence of simplicity after Borel resummation.\footnote{The tameness of Borel resummed asymptotic series in the context of QFT was studied in more detail in \cite{Grimm:2024hdx}.} 

Starting from a sufficiently simple QFT, e.g.~one whose Lagrangian has a finite complexity, it is a natural expectation that functions arising as physical observables have a limited information content, despite the growth of complexity order-by-order in perturbation theory. This leads us to conjecture that emergence of simplicity is a general phenomenon of these QFTs, or more precisely, that for a QFT with finite complexity, the exact observables can be represented with finite complexity.

\paragraph{Exact observables -- bootstrapping complexity.} A substantial part of the space of QFTs consists of theories for which no microscopic description exists, and in this case one can only rely on general physical principles such as unitarity, locality, and symmetries to constrain observables. What can we say about the complexity of exact observables in this case; in other words, can we bootstrap complexity?

As an example, consider two- and three-point functions of primary fields in a CFT. Conformal symmetry is sufficient to constrain these to take the exact general form 
\begin{align}
    \expval{\cO_1(x_1)\cO_2(x_2)} &= \frac{c_{12}\,\delta_{\Delta_1,\Delta_2}}{|x_1-x_2|^{\Delta_1+\Delta_2}} \,, \\
    \expval{\cO_1(x_1)\cO_2(x_2)\cO_3(x_3)} &= \frac{c_{123}}{|x_1-x_2|^{\Delta_1+\Delta_2-\Delta_3}|x_2-x_3|^{\Delta_2+\Delta_3-\Delta_1}|x_1-x_3|^{\Delta_1+\Delta_3-\Delta_2}}\,.
\end{align}
These functions admit a simple Pfaffian chain representation, so they have a finite complexity, solely based on conformal symmetry.

A powerful feature of our notion of functional complexity of observables is that it can be determined without having a closed-form expression. Instead, it is sufficient to find an implicit representation of the function, e.g.~by means of a Pfaffian differential equation. These differential equations can often be found by relying on underlying properties of the theory, thereby avoiding an explicit computation. As an example, it was recently shown that cosmological correlators satisfy a differential equation which can be derived algorithmically \cite{Arkani-Hamed:2023kig}, and this differential equation was shown to be Pfaffian in \cite{Grimm:2024mbw}.

Rather than using bootstrap methods to infer the complexity of an observable, one may also utilize complexity as a bootstrap principle. For example, we could impose a complexity bound on $n$-point correlation functions, i.e.~a maximal format and degree $\cF_n$ and $\cD_n$ depending on $n$, and study the space of QFTs satisfying these bounds without reference to a miscroscopic description. It would be interesting to explore this in future research, for instance by connecting it to Conjecture 5 of \cite{Douglas:2023fcg} on the space of CFTs satisfying bounds on the number of degrees of freedom.

\paragraph{Observables and complexity classes.} So far we have considered QFT observables as individual entities and analyzed their complexity as a function of kinematic variables and theory parameters. However, our initial goal, as formulated in question (ii), was to quantify the complexity of a QFT. In principle a QFT has observables of arbitrarily high complexity, e.g.~scattering amplitudes with many external particles or correlation functions with many operator insertions. It is then perhaps not reasonable to expect that the complete collections of observables in a QFT can be described with a finite amount of information. Instead, one approach is to quantify the growth of the complexity of observables as a function of some external label $N$, e.g.~the number of external particles or field insertions. The dependence on $N$ could then be classified by certain complexity classes, similar to computational complexity. This growth class could then be an indicator of the complexity of a QFT.

As a toy model example, consider again the $\phi^4$ theory on a point, whose observables are the correlation functions\footnote{In this theory, the observables depend only on the parameters of the theory since there are no kinematic variables.}
\begin{equation}\label{eq:phi4corr}
    I_N(g) = \int \dd \phi \, \phi^N \, e^{-S[\phi]} \,.
\end{equation}
These can be obtained non-perturbatively through a combination of recursion relations and differential equations \cite{Weinzierl:2022mmp}, and in \cite{Grimm:2023xqy} it was shown that $I_N$ has complexity $(\cF,\cD)=(4,8+\lceil N/4\rceil)$, growing stepwise linearly with $N$. Following the discussion above, one would then assign to this QFT the complexity class $(\cF,\cD)\sim \big(O(1),O(N)\big)$. Note that the growth of the format and degree have to be quantified separately. In this example, only $\cD$ grows, and since computational complexity grows polynomially in $\cD$, this indicates that there are algebraic relations present among the observables which reduce their information content. This example strengthens the idea that it is meaningful to quantify complexity by two numbers, as already noted in the mathematical literature.

\paragraph{Total information of observables.} Another approach to measuring the complexity of a QFT through its observables is to exploit algebraic relations among observables, as encountered in the example of the previous paragraph, to reduce the \textit{total} information content to a finite amount. With this in mind, we reformulate our question: is there a finite amount of data, from which all observables can be generated through a simple prescription? Here it is important to specify what we precisely mean by simple prescription; one could take an extreme viewpoint and consider a Lagrangian of finite complexity to be the finite set of initial data, and the quantization of the theory to be the prescription for obtaining the observables. Instead, we will require that the simple prescription consists of elementary logical steps, e.g.~an algebraic recursion relation.

In some simple settings, the answer to our question is known to be the affirmative. For example, consider a $d$-dimensional scalar QFT with polynomial interactions up to degree $D$ (i.e.~described by the Lagrangian $\cL_\text{poly}$ in equation \eqref{eq:Lpoly}), which is regularized by putting it on a finite lattice $\Lambda$. Enumerating the points on the lattice by $j=1,\ldots,|\Lambda|$, the correlation functions of this theory take the general form 
\begin{equation}
    I_{N_{1}\cdots N_{|\Lambda|}}(\lambda_1,\ldots,\lambda_D) = \int \dd\phi_1\cdots \dd\phi_{|\Lambda|} \, \phi_1^{N_1} \cdots \phi_{|\Lambda|}^{N_{|\Lambda|}} \, e^{-S[\phi]}\,.
\end{equation}
It was argued in \cite{Weinzierl:2020nhw} using twisted cohomology that all these correlation functions in this theory can be expressed as a linear combination of a finite set of basis integrals of size $(D-1)^{|\Lambda|}$, where $|\Lambda|$ is the number of points on the lattice. The coefficients of this linear combination may be found algorithmically by a reduction to master integrals, as explained in \cite{Weinzierl:2020nhw}.

In general, finding representations of lower complexity for observables, usually scattering amplitudes, is an active topic of research. It would be interesting for future research to attempt to make these methods quantitative, by measuring the functional complexity in terms of format and degree.

%%%%%%%%%%%%%%%%%%%%%%%%%%%%%%%%%%%%%%%%%%%%%%%%%%%%%%%%%%%%%%%%
\subsection{Action to observables}\label{sec:actiontoamp}
How is the complexity of the action of a theory related to the complexity of its observables? Having a measure of complexity for the action and observables of a QFT enables us to address this question quantitatively. We begin perturbatively by analyzing the complexity of Feynman diagrams in scalar theories.

\paragraph{Complexity of perturbation theory.}
A natural expectation is that theories with an action of higher complexity will have more complicated Feynman rules, and therefore a more complicated perturbation theory. For example, consider a scalar QFT with an algebraic scalar potential of complexity $(\cF,\cD)$. Every monomial in the potential yields a Feynman rule, and hence there are at most $(\cF+\cD)!/(\cF!\cD!)$ vertices in the theory. For potentials with only a few interactions, this estimate is sharpened by using the fewnomial representation.

The complexity of the perturbative expansion can be further probed by counting the number of diagrams $N_{\ell,n}$ for a fixed number of loops $\ell$ and external particles $n$. The growth of $N_{\ell,n}$ in $\ell$ and $n$ depends on the format and degree $(\cF,\cD)$ of the Lagrangian. For algebraic Lagrangians, it is clear that increasing $\cF$ and $\cD$ yields a significantly faster growth of $N_{\ell,n}$.

Suppose now that we have a QFT with an analytic potential of finite complexity $(\cF,\cD)$. The potential then generically has infinitely many interaction terms when expanded as a power series, and perturbatively we therefore have infinitely many interactions. At any finite order in perturbation theory, the additional structure in the potential is hidden, only becoming visible non-perturbatively. Therefore, to fully capture how the complexity of the action relates to the complexity of the observables, we have to go beyond perturbation theory and consider exact observables.

\paragraph{Action to exact observables.} In the spirit of the emergence of simplicity discussed above, we expect the connection between the complexity of actions and exact observables to be clearer than the perturbative case. Interestingly, it is not always true that actions of high complexity yield observables of high complexity. Most notably, in \cite{Arkani-Hamed:2008owk} it was pointed out that, from the perspective of amplitudes, the simplest QFTs are those with a high amount of supersymmetry, such as $\cN=4$ super Yang-Mills and $\cN=8$ supergravity. The enormous symmetry groups of these theories result in miraculous simplifications in the amplitudes. On the other hand, the actions for these theories have a high complexity due to the complicated structure of supersymmetry.

%%%%%%%%%%%%%%%%%%%%%%%%%%%%%%%%%%%%%%%%%%%%%%%%%%%%%%%%%%%%%%%%

\subsection{Renormalization group flow} A fundamental question is whether a lowering of the cut-off energy scale leads to a loss of information. In other words: what is the fate of QFT complexity along RG flow? Results such as the $c$-theorem \cite{Zamolodchikov:1986gt} the $a$-theorem \cite{Komargodski:2011vj} indicate that RG flow is irreversible and therefore lead to a loss of information as measured by the central charge $c$ or anomaly coefficient $a$. In this section we study what happens to format and degree $(\cF,\cD)$, which is another measure of information of a QFT, along RG flow.

\paragraph{Integrating out heavy fields.} To start our discussion, consider a theory with several fields, one of which is considered a heavy field in the sense that its mass $m$ is much greater than the cut-off scale $\Lambda$. RG flow is then often implemented by integrating out the heavy field, thereby obtaining an effective action for the remaining fields. This procedure reduces the number of degrees of freedom explicitly, at the expense of having more complicated interactions among the remaining fields. It will therefore be interesting to analyze what happens to $(\cF,\cD)$ when performing an integration over heavy fields.

In practice, one usually proceeds perturbatively and includes only finite loop corrections from integrating out the heavy field. In this approach, the complexity of the effective action will depend on how many loops $\ell$ are included. In particular, with our previous discussion in mind, we see that the complexity of the effective action will grow arbitrarily high by increasing $\ell$. To fully appreciate how complexity changes by integrating out a field, we therefore have to continue non-perturbatively and integrate out the field exactly. This clearly makes a general analysis a challenging task, and one can either proceed by considering the equations governing exact RG flow \cite{Rosten:2010vm}, or by explicitly analyzing tractable examples.

Here, we will only study a simple toy model to get an intuition for what happens to complexity under integrating out a field. Consider a 0d QFT on a point with two scalars $\phi_1$ and $\phi_2$, described by the action 
\begin{equation}
  S[\phi_1,\phi_2] = \frac{1}{2}m_1^2\phi_1^2  + \frac{1}{2}m_2^2\phi_2^2 + \lambda (\phi_1^2+\phi_2^2 )^2 \,.
\end{equation}
The complexity of this theory is given by $(\cF,\cD)= (2,4)$. We assume that $m_1\ll m_2$ now integrate out the heavy field $\phi_2$. This should lead to an effective action $S_{\rm eff}[\phi_1]$ which preserves the partition function, i.e.
\begin{equation}
   \int \dd \phi_1 \, e^{-S_{\rm eff}[\phi_1]} =  \int \dd\phi_1 \dd\phi_2 \, e^{-S[\phi_1,\phi_2]} \,.
\end{equation}
Since the path integral in this theory is exactly solvable (cf.~the discussion in section \ref{sec:amp}), we can evaluate the exact effective action and find
\begin{equation}
    S_{\rm eff}[\phi_1] = \frac{m_1^2}{2}\phi_1^2 + \lambda \phi_1^4 - \log\left[\frac{\sqrt{m_2^2+4\phi_1^2}}{2\sqrt{2 \lambda}} e^{\frac{(m_2^2+4\phi_1^2)^2}{32\lambda}} K_{1/4} \left( 
  \frac{(m_2^2+4\phi_1^2)^2}{32\lambda}\right) \right]  \,.
\end{equation}
The functions appearing in the effective Lagrangian can be written in terms of a Pfaffian chain \cite{Grimm:2023xqy}, and the format and degree of the resulting theory is significantly higher than the original theory. In particular, although there are fewer variables contributing to the format, the amount of auxiliary functions needed to generate the effective dynamics for $\phi_1$ still leads to a higher value of $\cF$.

Although this is a simple example, we expect that this is a general phenomenon: integrating out a field exactly will generate complicated functions for the effective action. This generically leads to an increase in degree, but the change in format is more subtle: there is a trade-off between the complexity loss of the integrated out fields and interactions of the UV Lagrangian, and the complexity gain due to the necessary functions needed for the IR Lagrangian. This situation may become particularly interesting upon integrating out a $N$ fields, or even an infinite tower of fields, since this may lead to an emergence of simplicity. Emergence of this type is known, for example, from the Schwinger one-loop computations of \cite{Gopakumar:1998ii,Gopakumar:1998jq} and is consistent with the more courageous emergence proposal put forward in \cite{Grimm:2018ohb,Heidenreich:2018kpg,Palti:2019pca}.

The above discussion indicates that the complexity of a QFT action under RG flow likely has interesting properties which warrant further study.
In particular, it is a challenging task to obtain a full picture of the behavior of complexity under renormalization, including possible emergence effects. In contrast, from the point of view of the observables, it is clear that the complexity of the total set of observables is reduced, since the UV observables are no longer part of the spectrum.

%%%%%%%%%%%%%%%%%%%%%%%%%%%%%%%%%%%%%%%%%%%%%%%%%%%%%%%%
\section{Conclusions and outlook}\label{sec:concl}

In this work we have shown that in order to quantify the complexity of a QFT, whether one starts from the action or the observables, one is naturally led to study the mathematical problem of assigning a well-defined complexity to functions and domains. Following recent mathematical developments, we have argued that the notion of format and degree $(\cF,\cD)$ from sharp o-minimality provides a meaningful solution to this problem, and we have explored how to implement this for QFTs. We began by assigning a format and degree to the Lagrangian of a QFT, and we found that $\cF$ and $\cD$ encode the number of degrees of freedom as well as the complexity of the interactions. Remarkably, our approach is valid even when the Lagrangian superficially seems to contain an infinite amount of information, e.g.~in the form of infinitely many interaction terms. This was accomplished by finding a different way of representing the functions and domains needed to define the Lagrangian, for which only a finite amount of information is required. For example, we have shown that solutions to certain differential equations have local and global information that can be measures with a complexity $(\cF,\cD)$. These insights clarify which QFTs have a finite complexity, and how this complexity can be quantified.

We then changed our perspective, focusing on the observables of a QFT rather than its microscopic description. From the outset, it is already a remarkable assertion that QFT observables have a finite information content. A first step towards this idea was taken in \cite{Douglas:2022ynw,Douglas:2023fcg}, where it was argued that QFT observables, under mild restrictions, fit into a class of functions which satisfies strong finiteness properties and avoids pathologies, formalized by o-minimality. In this work we taken a next step, by making this idea quantitative and giving a method of assigning a complexity to observables. Although our analysis is only in its infancy, we made some elementary observations on the behavior of complexity under various standard QFT aspects. We have seen that the complexity of an observable grows order-by-order in perturbation theory, but may exhibit a non-trivial emergence of simplicity in going towards the exact observable. This emergence of simplicity arises due to the presence of infinitely many terms, conspiring through algebraic relations, to form a function which can be described with finite information.

In addition to individual observables, we have also analyzed the complexity of a QFT by considering the set of all its observables. Here algebraic relations among observables played a key role in determining the information content of the set of observables. The complexity of observables can grow arbitrarily large by increasing external parameters such as the number of field insertions. We therefore quantified the complexity of this set by considering complexity classes which capture the growth of $(\cF,\cD)$ with these parameters. For example, we found that exact $N$-point correlation functions in 0d $\phi^4$ theory are described by the complexity class $\big(O(1),O(N)\big)$, which quantifies that the theory indeed has simple observables that can all be expressed with finitely many functions. In general, this notion of complexity classes may yield a promising approach for uncovering new structures in the space of observables of a QFT.

\subsubsection*{Outlook}
In our discussion, many questions emerged and there is much which remains to be investigated further. Below we highlight some of the most interesting avenues for future research.

\paragraph{Complexity of amplitudes.} While we have only given a short discussion of the complexity of amplitudes, we believe that it could be a valuable tool for gaining a deeper understanding of the structure of amplitudes. In particular, the conjecture of \cite{binyamini2024lognoetherianfunctions} on the sharp o-minimality of log-Noetherian functions implies that one can assign a complexity to all amplitudes in sufficiently tame QFTs.  
This vastly extends the scope of our proposal and may make complexity a widely applicable tool for QFT amplitudes. Concretely, the complexity of amplitudes could for instance be used to detect algebraic relations, find minimal representations, and compare QFTs through the complexity of their observables.

\paragraph{Quantum computational complexity and holography.}
In the context of QFT, there is a more widely studied notion of complexity which counts the quantum computational complexity of a state in the Hilbert space of the theory \cite{Jefferson:2017sdb,Chagnet:2021uvi}. Having its origins in quantum information theory, this measure of complexity plays a significant role in holography and is believed to be dual to the volume of a certain spatial slice in the bulk theory \cite{Susskind:2014rva,Brown:2015bva}. While no precise relation between the two notions of complexity has been established so far, a number of similarities were pointed out in \cite{Grimm:2023xqy}. Since format and degree may in principle be assigned to any sufficiently tame mathematical object, it would be fascinating to gain a deeper understanding of the connection between our notion of complexity and quantum computational complexity, and to investigate whether the complexity of a QFT has any holographic interpretation.

\paragraph{Complexity and entropy.} Sharp o-minimality gives a measure of complexity for functions and domains, and it is intriguing to apply these ideas to field configurations supported in a spacetime region, and thereby connect to physical limitations on information storage. This would entail that one establishes a bound on the maximum complexity which a field configuration is allowed to have within a spacetime region. Such a bound might be motivated by connecting complexity and various notions of entropy within a QFT and considering the Bekenstein bound. With such a complexity bound at hand, one could then identify states of maximum complexity. In view of the Bekenstein bound, these could be black holes, or more general maximum entropy objects such as the saturons considered in \cite{Dvali:2020wqi,Dvali:2021tez}. The fact that the construction of saturons appears to rely on a large $N$ limit suggests that they may be connected to the complexity of the underlying QFT.

\paragraph{Complexity and the swampland.} The swampland program aims to find what distinguishes EFTs coupled to gravity which admit a UV-completion to quantum gravity from those which do not. These ideas are often centered around finiteness properties of EFTs, in terms of e.g.~number of degrees of freedom, the rank of the gauge group, the vacuum structure, and amplitudes \cite{Vafa:2005ui,Hamada:2021yxy}. It was conjectured that one of the finiteness principles which these EFTs should satisfy is that the functions appearing in the effective Lagrangians are tame \cite{Grimm:2021vpn}. The refined notion of tameness discussed in this work therefore allows one to in principle assign a complexity to these EFTs. This opens the opportunity to ask fascinating questions about the quantum gravity landscape. Can the swampland conjectures be formulated using complexity? Can QFTs with infinite complexity be coupled to quantum gravity, or is there a species scale mechanism which forbids this? What is the most complex QFT which can arise from quantum gravity? We intend to address these questions in an upcoming work \cite{GV1}.

\subsubsection*{Acknowledgements}
We are immensely grateful to Gal Binyamini for sharing his deep understanding of the relevant mathematical concepts with us.
Furthermore, we would like to thank C\'esar Ag\'on, Mike Douglas, Gia Dvali, Arno Hoefnagels, Ro Jefferson, Jeroen Monnee, Eran Palti, David Prieto, and Javier Subils for useful discussions and comments. Furthermore, we thank the referees from JHEP for valuable comments on an earlier version. This research is supported, in part, by the Dutch Research Council (NWO) via a Vici grant.

%%%%%%%%%%%%%%%%%%%%%%%%%%%%%%%%%%%%%%%%%%%%%%%%%%%%%%%%%%%%%%%
%%%%%%%%%%%%%%%%%%%%%%%%%%%%%%%%%%%%%%%%%%%%%%%%%%%%%%%%%%%%%%%
%%%%%%%%%%%%%%%%%%%%%%%%%%%%%%%%%%%%%%%%%%%%%%%%%%%%%%%%%%%%%%%
%%%%%%%%%%%%%%%%%%%%%%%%%%%%%%%%%%%%%%%%%%%%%%%%%%%%%%%%%%%%%%%
%%%%%%%%%%%%%%%%%%%%%%%%%%%%%%%%%%%%%%%%%%%%%%%%%%%%%%%%%%%%%%%

\appendix

\section{Tameness and geometric complexity}\label{app:geom}
In this appendix we introduce the technical background material which underlies the discussion in the main text. We start with a brief introduction to the tameness principle implemented in o-minimal structures. We then discuss how these can be refined to sharply o-minimal structures, which have a notion of complexity. 

\subsection{O-minimality}
Tame geometry concerns itself with geometric objects which can be described in terms of a finite amount of information, and tries to find geometric structures which are as general as possible while maintaining essential finiteness features. A concrete tameness principle is implemented via the central object of an \textit{o-minimal structure}. This is a collection of subsets $\cS=(\cS_n)$ of $\bbR^n$ for every $n$, satisfying the following conditions:
\begin{enumerate}
    \item[(i)] $\cS$ is closed under finite unions, finite intersections, complements, products, and linear projections;
    \item[(ii)] $\cS$ contains at least all algebraic sets;
    \item[(iii)] every set in $\cS_1$ has a finite number of connected components.
\end{enumerate}
In other words, an o-minimal structure specifies which subsets of Euclidean space $\bbR^n$ we would like to allow in our geometry, subject to some consistency rules.  Sets which are contained in an o-minimal structure $\cS$ are called definable. The final axiom is the one that implements the idea of tameness, and its interaction with the other axioms ensure that every definable set satisfies a rich collection of tameness theorems \cite{VdDries}. While definability initially only refers to sets, it naturally allows us to extend the notion of definability for a function $f:X\to Y$ as well, by requiring that its graph $\Gamma(f)=\{(x,y)\in X\times Y\,|\, y=f(x)\}\subseteq X\times Y$ is a definable set. In this way, definable functions inherit all tameness features from definable sets.

\paragraph{Examples of o-minimal structures.} One of the main goals of tame geometry is to find the most general possible collections of functions and sets which satisfy these axioms. At present, there are many known o-minimal structures. Some of the significant examples which are relevant in this work are the following. 
\begin{enumerate}
    \item[-] $\bbR_{\rm alg}$. This structure is defined using polynomials, and its building blocks are simple semi-algebraic sets of the form 
        \begin{equation}
        X = \{x \,|\,P(x)=0, \, Q(x)>0\},
        \end{equation}
        where $P$ and $Q$ are polynomials. This is the smallest possible o-minimal structure, and it is already a non-trivial statement that these sets satisfy the axioms given above \cite{Tarski}. 
    \item[-] $\bbR_{\rm an,exp}$. This structure is generated by the restricted analytic functions and the exponential function $\exp:\bbR\to\bbR$; here restricted analytic functions are defined as restrictions of analytic functions to closed balls \cite{VdDriesMiller}. This structure is remarkably rich and captures many non-trivial geometric objects, such as those arising in asymptotic Hodge theory. 
    \item[-] $\bbR_{\rm Pfaff}$. This structure is generated by Pfaffian functions as defined in the main text. This o-minimal structure is known for its tractable notion of complexity, which we already used implicitly in our first proposal of QFT complexity. If one demands that the domains of the Pfaffian functions are compact and analytic on their domain, one obtains the smaller o-minimal structure $\bbR_{\rm rPfaff}$. 
\end{enumerate}

In recent works it has been argued that certain physical systems exhibit tameness, in the sense the corresponding physical quantities should be definable within an o-minimal structure. These ideas started in the setting of EFTs which admit a UV-completion to quantum gravity \cite{Grimm:2021vpn}, and were later investigated for general QFTs \cite{Douglas:2022ynw,Douglas:2023fcg,Grimm:2023xqy,Grimm:2024mbw,Grimm:2024hdx}.

While o-minimality is an elegant mathematical principle, a fundamental shortcoming is that it is only a qualitative attribute. More explicitly, any countable topological or geometric feature\footnote{Examples of such features are the number of connected components, dimensions of homology or cohmology groups, number of discontinuities, number of simplices in a simplicial complex representation, et cetera.} of a definable set is finite, but the framework does not tell us how finite. Recently, efforts have been made to find a quantitative version of tameness, which is \textit{effective} in the sense that these finite numbers can be explicitly computed \cite{binyamini2022sharply,beyondo-min,binyamini2024lognoetherianfunctions}. This brings us to the introduction of a sharpened notion of tameness going beyond o-minimality.

\subsection{Sharp o-minimality}
To quantitatively control the finiteness implemented in an o-minimal structure, a proposal was made to define a notion of complexity for definable sets \cite{binyamini2022sharply,beyondo-min,binyamini2024lognoetherianfunctions}. The fact that o-minimality is an extremely general framework makes this a challenging task, and in order to achieve this, we must endow an o-minimal structure with a sensible notion of `complexity measure'. Given a definable set $X$, the basic idea is to write it in terms of some logical formulas, and then to count the information contained in these formulas. In order to be a sensible measure, this counting must obey some basic rules which dictate how complexity behaves under geometric operations.

Let us now discuss the precise construction. The complexity in an o-minimal structure $\cS$ is measured as follows: we organize the sets in $\cS$ into collections $\Omega_{\cF,\cD}$, where $\cF,\cD\in \mathbb{N}$ are two integers serving as indices. These collections are assumed to form a filtration, meaning that
\begin{equation}
    \Omega_{\cF,\cD} \subseteq \Omega_{\cF,\cD+1}, \quad \Omega_{\cF,\cD} \subseteq \Omega_{\cF+1,\cD} 
    \quad \text{and}\quad  
    \bigcup_{\cF,\cD}\Omega_{\cF,\cD}=\cS  \, . 
\end{equation}
The integers $\cF$ and $\cD$ indexing the filtration are called format and degree, respectively. The filtration $\{\Omega_{\cF,\cD}\}$ is required to satisfy the following conditions: 
\begin{enumerate}
    \item[(i)] Complements and linear projections preserve format and degree, and products with the real line increase the format by one: if $A\subseteq \bbR^n$ and $A\in \Omega_{\cF,\cD}$, then 
    \begin{equation}
        \bbR^n \backslash A,\, \pi(A) \in \Omega_{\cF,\cD},
    \end{equation}
    with $\pi:\bbR^n\to \bbR^{n-1}$ a linear projection, and
    \begin{equation}
        A\times \bbR, \, \bbR\times A \in \Omega_{\cF+1,\cD}\,.
    \end{equation}
    \item[(ii)] When taking unions or intersections, the format is maximized and the degrees are added: if $A_i\subseteq \bbR^n$ and $A_i\in \Omega_{\cF_i,\cD_i}$ for $i=1,\ldots,k$, then
    \begin{equation}
        \bigcup_{i=1}^k A_i, \, \bigcap_{i=1}^k A_i \in \Omega_{\cF,\cD}\,,
    \end{equation}
    where $\cF=\max(\cF_1,\ldots,\cF_k)$ and $\cD=\cD_1+\cdots+\cD_k$.
    \item[(iii)] The format is at least the ambient dimension of a set: if $A\subseteq \bbR^n$ and $A\in \Omega_{\cF,\cD}$, then $\cF\geq n$;
    \item[(iv)] The complexity of algebraic sets is fixed by the number of variables and the degree: if $P(x_1,\ldots,x_n)$ is a polynomial, then
    \begin{equation}\label{eq:algFD}
        \{(x_1,\ldots,x_n)\in \bbR^n \,|\, P(x_1,\ldots,x_n)=0\} \in \Omega_{n,\deg P} \,.
    \end{equation}
    \item[(v)] For every format $\cF$ there is a fixed polynomial $P_\cF$ with positive coefficients, such that whenever $A\subseteq \bbR$ and $A \in \Omega_{\cF,\cD}$, $A$ has at most $P_\cF(\cD)$ connected components. 
\end{enumerate}
The first two axioms tell us how format and degree behave under geometric operations. Axioms (iii) and (iv) then ensure that $\cF$ encodes the number of variables, and that $\cD$ reduces to the familiar notion of degree in the algebraic case. Finally, axiom (v) is the quantitative version of axiom (iii) of o-minimality, and it results in a sharpened notion of tameness. An o-minimal structure together with a complexity filtration $\{\Omega_{\cF,\cD}\}$ is called a \textit{sharply o-minimal structure} \cite{binyamini2022sharply,beyondo-min,binyamini2024lognoetherianfunctions}.

At first glance the appearance of an infinite collection of polynomials $P_\cF$ seems rather abstract and unpractical, but it is implicitly understood in this definition that these polynomials can be calculated explicitly. The purpose of these polynomials is to provide sharp estimates for any finiteness feature associated to definable sets. The definition of o-minimality is such that finiteness always reduces to finiteness of projected one-dimensional connected components, which are controlled by $P_\cF(\cD)$. In this way, the axioms guarantee that the finiteness features of tame sets can be quantified by a polynomial in $\cD$, depending on $\cF$.

Note that the complexity filtration $\{\Omega_{\cF,\cD}\}$ is a choice of additional data on top of an o-minimal structure. While this may initially appear to be a disadvantage of the framework, it is in fact a common feature of complexity. For instance, the notion of quantum computational complexity inspired by Nielsen's geometric approach \cite{nielsen2005geometricapproachquantumcircuit,Jefferson:2017sdb} relies on a choice of complexity measure on the space of unitary operators.\footnote{More precisely, this construction requires a choice of Finsler metric which serves as a cost function.} Additionally, let us note that there is a notion of equivalence between complexity filtrations which can be used to tell whether the resulting complexity is comparable \cite{binyamini2022sharply,beyondo-min}.

\paragraph{Examples of sharply o-minimal structures.} Let us now list some examples, conjectural examples, and non-examples of sharply o-minimal structures.
\begin{enumerate}
    \item[-] $\bbR_{\rm alg}$. This is the most natural example of a sharply o-minimal structure, since the notion of degree can be directly inferred from the underlying algebraicity, using the condition of equation \eqref{eq:algFD}
    satisfied by algebraic hypersurfaces. It was shown in \cite{binyamini2022sharply} that the complexity filtration generated by this condition satisfies the axioms of sharp o-minimality. Note that a polynomial $P(x_1,\ldots,x_n)$ then has complexity $(n+1,\deg P)$, since its graph is the zero set of the polynomial $Q(y,x_1,\ldots,x_n)=y-P(x_1,\ldots,x_n)$ which has $n+1$ variables. 
    \item[-] $\bbR_{\rm an,exp}$. This structure is \textit{not} sharply o-minimal, as noted in \cite{beyondo-min}. The reason is that this o-minimal structure contains the class of restricted analytic functions, which are too general to admit any polynomial complexity bounds, as required by sharp o-minimality. An example of a function definable in $\bbR_{\rm an,exp}$ which does not admit such bounds is the function 
    \[
    f(z) = \sum_{k=1}^\infty z^{\omega_k}
    \]
    defined on the disk of radius $1/2$, where the sequence $\omega_k$ is inductively defined by $\omega_k=2^{\omega_{k-1}}$ and $\omega_1=1$.
    \item[-] $\bbR_{\rm rPfaff}$. Using the strong finiteness theorems obeyed by sets constructed from Pfaffian functions, it has been shown that this o-minimal structure admits a complexity filtration satisfying the axioms listed above \cite{EffCell}. For the details of the construction we refer to the original work and \cite{beyondo-min}. The complexity filtration is such that semi-Pfaffian sets satisfy
    \begin{equation}
        \{(x_1,\ldots,x_n)\in \bbR^n \,|\, f(x_1,\ldots,x_n)=0,\, g(x_1,\ldots,x_n) >0\} \in \Omega_{\cF,\cD} \,,
    \end{equation}
    where $\cF=n$ and $\cD$ is the sum of the degrees of the polynomials required in the construction of $f$ and $g$ as Pfaffian functions. To assign a complexity to all sets in the structure, one has to address some subtleties involving complements \cite{beyondo-min}.
    It is conjectured that the structure $\bbR_{\rm Pfaff}$ which contains unrestricted Pfaffian functions is also sharply o-minimal \cite{beyondo-min}.
    \item[-] $\bbR_{\rm LN,PF}$. This o-minimal structure was introduced recently in \cite{binyamini2024lognoetherianfunctions}, and is generated by the Pfaffian extension of log-Noetherian functions as explained in section \ref{sec:generalframework}. While it is conjectured that this structure is sharply o-minimal, it has been shown that this structure admits a notion of complexity which is weaker than sharp o-minimality, called effective o-minimality.\footnote{The main difference is that, instead of having a format and degree, objects in this structure only have a format. The format controls explicit geometric and computational complexity bounds, but can grow arbitrarily fast, unlike the polynomial growth in degree for sharply o-minimal structures.} This is significant progress in proving the conjecture that periods arising in Hodge theory admit a notion of complexity in the form of sharp o-minimality \cite{beyondo-min}. 
    \end{enumerate}

Let us mention some important aspects of sharp o-minimality. First, we note that the complexity filtration $\{\Omega_{\cF.\cD}\}$ is not merely an organizational tool for tame sets in an o-minimal structure, but actually provides quantitative finiteness statements, encoded in the polynomials $P_\cF$. For example, one of the crucial theorems for o-minimal structures is the cell decomposition theorem, which essentially states that tame sets can be constructed as the union of finitely many simple sets called cells. For sharp o-minimality, this mere finiteness statement is sharpened by bounding the number of cells and the complexity of the cells polynomially in terms of $\cF$ and $\cD$ \cite{beyondo-min}. The same principle can be applied to any finiteness theorem in o-minimality. 

Finally, note that for a given definable set $A$ the format and degree are not unique; there may be several pairs $(\cF,\cD)$ for which $A\in\Omega_{\cF,\cD}$. This uniqueness challenge was discussed in \cite{Grimm:2023xqy}, where it was suggested that the \textit{sharp complexity} of $A$ should consist of the finitely many pairs $(\cF,\cD)$ which are minimal in the sense that they satisfy the condition 
\begin{equation}
A \in \Omega_{\cF,\cD}, \quad  A \notin \Omega_{\cF,\cD-1},\quad  A \notin \Omega_{\cF-1,\cD}\,.
\end{equation}
An essential observation is that the `minimal complexity' $(\cF,\cD)$ depends on the problem at hand, since different pairs may yield sharper bounds in different computational scenarios. In our applications to QFT, there is usually a natural choice for $(\cF,\cD)$.

\bibliography{main}
\bibliographystyle{utphys}

\end{document}